\documentclass[12pt]{article}

\usepackage{amssymb}

\begin{document}

\begin{titlepage}

\begin{flushright}
arXiv:1110.2123
\end{flushright}
\vskip 2.5cm

\begin{center}
{\Large \bf Consequences of Neutrino Lorentz Violation\\
For Leptonic Meson Decays}
\end{center}

\vspace{1ex}

\begin{center}
{\large Brett Altschul\footnote{{\tt baltschu@physics.sc.edu}}}

\vspace{5mm}
{\sl Department of Physics and Astronomy} \\
{\sl University of South Carolina} \\
{\sl Columbia, SC 29208} \\
\end{center}

\vspace{2.5ex}

\medskip

\centerline {\bf Abstract}

\bigskip

If the observation by OPERA of apparently superluminal neutrinos is correct, the
Lagrangian for second-generation leptons must break Lorentz invariance. We calculate
the effects of an energy-independent change in the neutrino speed on another
observable, the charged pion decay rate. The rate decreases by a factor
$\left[1-\frac{3}{1-m_{\mu}^{2}/m_{\pi}^{2}}
\left(\langle v_{\nu}\rangle-1\right)\right]$, where $\langle v_{\nu}\rangle$ is
the (directionally averaged) neutrino speed in the pion's rest frame. This provides
a completely independent experimental observable that is sensitive to the same
forms of Lorentz violation as a neutrino time of flight measurement.

\bigskip

\end{titlepage}

\newpage

The recent announcement by the OPERA experiment that muon neutrinos may
be traveling faster than light with speeds of $\sim1+(2.5\times10^{-5})$~\cite{ref-opera}
has attracted a tremendous amount of attention. Whether this experimental conclusion will withstand further scrutiny remains
to be seen. However, the mere possibility that superluminal travel might occur has
already produced an explosion of interest in theories that do not precisely obey
Lorentz symmetry.

Before the intriguing neutrino results were reported, there was already a
thriving community of physicists working on such Lorentz-violating theories. This
community includes both experimenters and theorists. One of the most important
theoretical developments in the field was an effective quantum field theory capable
of describing all possible forms of Lorentz violation that might occur at low energies
with standard model fields~\cite{ref-kost1,ref-kost2}.
This theory is known as the standard model extension (SME).
Both the renormalizability~\cite{ref-kost4,ref-colladay2,ref-ferrero1} and
stability~\cite{ref-kost3} of the SME have been studied.
The SME is now the standard framework for parameterizing the results of experimental
Lorentz tests. Many different systems and phenomena may be used for
such tests, and up-to-date information about
most of the resulting constraints may be found in~\cite{ref-tables}.

In this paper, we shall use the SME formalism to examine one very important
consequence that the existence of superluminal neutrinos could have. If the
neutrino energy-momentum relation is modified, the rates of scattering and decay
processes involving these neutrinos will necessarily be modified. Of particular
interest are meson decays. The phase space available to the decay products in a
process such as $\pi^{+}\rightarrow\mu^{+}+\nu_{\mu}$ depends crucially on the
neutrino velocity. This provides a completely new observable, which is nonetheless
sensitive to some of the same parameters as the OPERA experiment. Precision decay
measurements can thus be directly compared with the results of a neutrino time of
flight experiment.

We shall calculate how an energy-independent change in the muon neutrino speed
affects the decay rate of a charged pion. The choice of the pion is partially just for
definiteness. However, we shall also find that the Lorentz-violating modifications to
the decay rate are most pronounced for the lightest parent particles. While this makes
pions the ideal species to study, our calculations naturally apply to any particle
that decays via a $W$ boson primarily to a muon and its associated neutrino.

Moreover, we expect that similar phenomena would occur in more complicated
decay processes, such as $\mu^{+}\rightarrow e^{+}+\nu_{e}+\bar{\nu}_{\mu}$.
In fact, calculations of Cerenkov-like decay processes such as
$\nu_{\mu}\rightarrow\nu_{\mu}+e^{+}+e^{-}$ have already raised
serious questions about
the OPERA result. Such processes, which could be allowed for superluminal
neutrinos, would have major effects on the neutrino beam in transit---effects which
have seemingly not been observed~\cite{ref-bi,ref-cohen}.

In our calculations,
we shall assume that the only Lorentz violation is in the lepton sector. The 
structures and interactions of the pion itself and the $W$ that mediates the decay
are exactly as in the standard model. This means that the entire calculation may
be performed, as usual, in the rest frame of the pion. The kinematic factors
associated with the motion of the parent particle are unchanged from the conventional
case.

However, it is not possible in the SME formalism to have Lorentz violation for
muon neutrinos but simultaneously to have
a muon sector that is
free of Lorentz violation. The lepton Lagrange density that contains a spin- and
energy-independent modification to the neutrino velocity is
\begin{equation}
{\cal L}=i\bar{L}(\gamma^{\mu}+c^{\nu\mu}\gamma_{\nu})D_{\mu}L+
i\bar{R}(\gamma^{\mu}+c^{\nu\mu}\gamma_{\nu})D_{\mu}R\,+
{\rm mass\,terms},
\end{equation}
where $L$ and $R$ are the left- and right-chiral lepton multiplets,
\begin{equation}
L=\left[
\begin{array}{c}
\nu_{L} \\
\ell_{L}
\end{array}
\right],\,\,\,
R=\left[\ell_{R}\right].
\end{equation}
and $D_{\mu}$ is the covariant derivative, containing the electromagnetic and
weak gauge interactions.
$SU(2)_{L}$ gauge invariance requires that the Lorentz violation coefficients
$c^{\nu\mu}$ be the same for the charged lepton and its neutrino.
(Actually, it is possible to have a different set of coefficients for the
right-chiral charged lepton field, but those coefficients are irrelevant in
reactions involving charged current weak interactions.) The $c$ coefficients
together form a
traceless background tensor, which imbues spacetime with a preferred
frame structure. While the $c$ coefficients control violations both of isotropy and
of boost invariance, we shall mostly be concerned here with the isotropic
element of the tensor $c_{00}$. Because Lorentz violation is a small effect, we may
neglect all terms beyond first order in $c$.

The choice of an energy-independent modification of the lepton velocities is motivated
by two important factors. First, the OPERA data does not show evidence of any energy
dependence in the neutrino speed. Second, the observed effect is rather large,
suggesting that the modification of the velocity is not an effect suppressed by any
powers of a large energy scale. This requires that the Lorentz-violating operator
responsible for the effect be renormalizable, leading to the choice of $c$.

There are very few direct constraints on the $c$ coefficients for second-generation
leptons. The absence of the photon decay process $\gamma\rightarrow\mu^{+}+\mu^{-}$
up to TeV photon energies constrains the possibility that the leptons might have
limiting velocities less than the speed of light~\cite{ref-altschul14},
but there are few
constraints on the possibility that muons and their corresponding neutrinos might
move faster than light. Superluminal muons would emit vacuum Cerenkov
radiation~\cite{ref-altschul9}, but to constrain the muon $c$ coefficients at the
level seen by OPERA would require careful observation
of $\sim 15$ GeV muons.

The strongest bounds on other forms of muon Lorentz violation are based on
$g_{\mu}-2$ measurements~\cite{ref-bennett}. This data could also be used to
constrain the muon $c$, potentially at the $10^{-6}$ level.
The behavior of a fermion in an external field, with a form of photon-sector Lorentz
violation that is essentially equivalent to $c$, has been calculated. The result is
that the spin
couples to both the electric and magnetic fields in a potentially anisotropic
fashion~\cite{ref-carone}. A full analysis of this phenomenon
would need to include the effects of
the substantial $\vec{E}$ fields present in muon $g_{\mu}-2$ experiments
as well as the possible influences of other SME parameters.

On the other hand, there are some constraints on $c$-type Lorentz violation in the
neutrino sector. If the $c$ coefficients are nondiagonal in flavor space, they will
lead to neutrino oscillations, with a different characteristic energy dependence
from the oscillations generated by nondiagonal neutrino
masses~\cite{ref-kost22}. This possibility has been constrained, and Lorentz
violation as the principal cause of observed neutrino oscillations has already been
ruled out~\cite{ref-barger1}. Oscillation measurements can also constrain the
differences between the flavor-diagonal $c$ parameters for different species, since
if the different neutrino flavors had substantially different speeds, the wave packet
overlap necessary for oscillations could be destroyed. With only isotropic $c$
coefficients, the oscillation data require a very high degree of flavor independence
(at the $10^{-19}$ level), which would make the OPERA data incompatible with numerous
measurements of electron-sector Lorentz violation~\cite{ref-giudice,ref-drago}.
However, the number of
SME parameters that can influence neutrino oscillations is very large, and it is not
clear whether this conclusion applies in more general models.
Finally, there is the famous time of flight measurement made on neutrinos
from SN 1987A~\cite{ref-longo,ref-stodolsky}, which (assuming the interpretation of
the data was correct) constrains the $c$ coefficients for electron neutrinos at
the $10^{-9}$ level. 

Constraints on the SME parameters are conventionally expressed in a sun-centered
reference frame, with coordinates $(X,Y,Z,T)$, with the $Z$-direction parallel to the
Earth's rotation axis~\cite{ref-bluhm4}.
The OPERA neutrino timing data was collected over a long period of time, effectively
averaging over the range of neutrino directions $\hat{v}$ that are possible as the
Earth rotates. As shown below in (\ref{eq-v}), the speed of a massless particle
in the presence of the $c$ parameters is $1-c_{00}-c_{(0j)}\hat{v}_{j}-
c_{jk}\hat{v}_{j}\hat{v}_{k}$, where $c_{(0j)}$ is the symmetrized combination
$c_{0j}+c_{j0}$.
This makes the average speed of the OPERA neutrinos
$1-c_{00}-c_{(0Z)}\cos\theta-\frac{1}{2}(c_{XX}+c_{YY})\sin^{2}\theta-c_{ZZ}
\cos^{2}\theta$, where $\theta$ is the angle between the CERN-to-Gran Sasso direction
and the $Z$-axis, for which $\cos\theta\approx-0.41$.

We now turn to the calculation of the pion decay rate.
This rate is not directly sensitive to all the $c$ parameters that might have
affected the average neutrino speed.
Because, in the absence of Lorentz violation, the distribution of the decay
products is isotropic, the modifications to the total decay rate can depend only on
the isotropic $c_{00}=c_{jj}$, as measured in the decaying
particle's rest frame. We may therefore treat $c$ as if it were a
diagonal tensor,
with elements $[c_{00},\frac{1}{3}c_{00},\frac{1}{3}c_{00},\frac{1}{3}c_{00}]$ in
the center of mass frame. If this isotropic $c$
is the sole contributor to the neutrinos' faster-than-light
travel, $v_{\nu_{\mu}}=1-\frac{4}{3}c_{00}$. However, the value of $c_{00}$ will
depend on the pion momentum, and other components of $c$ will affect the
differences in the decay rates of pions with different velocities.

Although the ultimate form taken by the modified decay rate is quite simple,
the calculations leading up to it are somewhat tricky. Methods for calculating
cross sections and decay rates in the SME context have been worked
out~\cite{ref-kost5}. The formulas for physical observables can be put in forms
quite similar to those that appear conventionally. However, in these formulas,
both the matrix element ${\cal M}$ for the decay and the kinematics are modified from
their conventional forms. The kinematic modifications, which are related to
changes in the phase space available to the external particles, are often more
important than the changes to the matrix element. 

The matrix element ${\cal M}$ can be determined from the same tree-level Feynman
diagram that governs conventional pion decay. Since there is assumed to be no
Lorentz violation in the meson or $W$ boson sectors, only those parts of the diagram
involving the final muon and neutrino need to be considered. This means that
${\cal M}$ may be modified in two ways. The first is
via the replacement of
$\gamma^{\mu}$ at the muon-neutrino-gauge boson vertex with
$\gamma^{\mu}+c^{\mu\nu}\gamma_{\nu}$. The free index $\mu$ is ultimately
to be contracted with
the total momentum vector $p_{\mu}$, which has only a time component. This 
amounts to a rescaling of the matrix element simply by a factor of
$(1+c_{00})$.

The second modification of the matrix element comes from the modified spinors,
which obey the modified Dirac equation
\begin{equation}
\left[(1+c_{00})\gamma_{0}E-\left(1-\frac{1}{3}c_{00}\right)\gamma_{j}p_{j}-m\right]
u(p)=0.
\end{equation}
The spinor solutions are standard, except for an evident rescaling of $E$ and $p_{j}$.
Therefore, if they are normalized to obey $\bar{u}^{s'}(p)u^{s}(p)=2E\delta_{ss'}$
[and analogously for $v(p)$], the closure relations
needed for calculating cross sections are
$\sum_{s}u^{s}(p)\bar{u}^{s}(p)=\gamma^{0}E+\left(1-\frac{4}{3}c_{00}\right)
\gamma_{j}p_{j}+m$ and
$\sum_{s}v^{s}(p)\bar{v}^{s}(p)=\gamma^{0}E+\left(1-\frac{4}{3}c_{00}\right)
\gamma_{j}p_{j}-m$.
This makes the matrix element squared proportional to
\begin{eqnarray}
|{\cal M}|^{2} & \propto & (1+c_{00})^{2}\sum_{s,s'}
{\rm tr}\,\left\{v_{\nu}^{s'}(-\vec{p}_{\mu})\gamma_{0}(1-\gamma_{5})
u_{\mu}^{s}(\vec{p}_{\mu})\bar{u}_{\mu}^{s}(\vec{p}_{\mu})(1+\gamma_{5})\gamma_{0}
v_{\nu}^{s}(-\vec{p}_{\mu})\right\} \\ 
& = & (1+c_{00})^{2}\,{\rm tr}\,\left\{
\left[\gamma_{0}E_{\nu}-\left(1-\frac{4}{3}c_{00}\right)\gamma_{j}(-p_{\mu})_{j}\right]\gamma_{0}(1-\gamma_{5})\right. \nonumber\\
& & \times\left.
\left[\gamma_{0}E_{\mu}-\left(1-\frac{4}{3}c_{00}\right)\gamma_{j}(p_{\mu})_{j}
\right](1+\gamma_{5})\gamma_{0}\right\} \\
\label{eq-M}
& =& 2\left(1+\frac{2}{3}c_{00}\right)p_{\mu}\left[E_{\mu}-\left(1-\frac{4}{3}c_{00}
\right)p_{\mu}\right].
\end{eqnarray}
In these formulas, $p_{\mu}$ and $p_{\nu}$ denote the magnitudes of the muon and
neutrino three-momenta. (Despite the indices, these symbols do not represent
four-vectors.) $E_{\mu}$ and $E_{\nu}$ are the corresponding energies.
The derivation of (\ref{eq-M}) used
the facts that the neutrino momentum $\vec{p}_{\nu}$ is opposite to the muon
momentum $\vec{p}_{\mu}$ and that $E_{\nu}=\left(1-\frac{4}{3}c_{00}\right)p_{\nu}$,
which is an obvious consequence of the massless Dirac equation.

Further simplification requires analysis of the more complicated dispersion relation
for the muon.
The energy of a fermion, including the leading order effects of $c$, is
\begin{equation}
E=\sqrt{(m^{2}+p_{j}p_{j})(1-2c_{00})-2c_{jk}p_{j}p_{k}-2c_{0j}p_{j}
\sqrt{m^{2}+p_{j}p_{j}}},
\end{equation}
which reduces to $E=\sqrt{(1-2c_{00})m^{2}+(1-\frac{8}{3}c_{00})p_{j}p_{j}}$
when only the isotropic $c_{00}$ is considered. Energy and momentum conservation
then dictate that the muon momentum is $p_{\mu}=\frac{1-\xi^{2}}{2}m_{\pi}\left[1+
\frac{2(2+\xi^{2})}{3(1-\xi^{2})}c_{00}\right]$, where $\xi=\frac{m_{\mu}}{m_{\pi}}$
is the muon-pion mass ratio---the main dimensionless parameter determining the
kinematics of the Lorentz-invariant case. The energies of the decay products are
$E_{\nu}=\frac{1-\xi^{2}}{2}m_{\pi}\left(1+\frac{2\xi^{2}}{1-\xi^{2}}
c_{00}\right)$ and $E_{\mu}=\frac{1+\xi^{2}}{2}m_{\pi}\left(1-\frac{2\xi^{2}}{1+\xi^{2}}
c_{00}\right)$. The quantity $E_{\mu}-\left(1-\frac{4}{3}c_{00}\right)p_{\mu}$ is
then $\xi^{2}m_{\pi}(1-2c_{00})$, so the ratio of the matrix element squared in
the presence of $c_{00}$ to its value at $c_{00}=0$ is
\begin{equation}
\left|\frac{{\cal M}}{{\cal M}_{0}}\right|^{2}=\left(1-\frac{4}{3}c_{00}\right)
\left(\frac{p_{\mu}}{p_{\mu 0}}\right),
\end{equation}
with the subscript 0 referring to the Lorentz-invariant values of particular
quantities.

As mentioned,
the Lorentz violation changes not just the matrix element for this process, but also
the phase space factors in the decay rate formula, since $c_{00}$ affects the free
propagation of the
daughter particles. The phase space integral appearing in the decay rate is
\begin{eqnarray}
\int d\Pi & = & \int\frac{d^{3}p_{\mu}}{(2\pi)^{3}}\frac{1}{2E_{\mu}}
\frac{d^{3}p_{\nu}}{(2\pi)^{3}}\frac{1}{2E_{\nu}}(2\pi)^{4}\delta^{4}
(p_{\pi}-p_{\mu}-p_{\nu}) \\
& = & \frac{1}{16\pi^{2}}\int d^{3}p_{\mu}\frac{1}{E_{\mu}(p_{\mu})E_{\nu}(p_{\mu})}
\delta(m_{\pi}-E_{\mu}-E_{\nu}).
\end{eqnarray}
After eliminating $p_{\nu}$,
it remains to evaluate the integral over the muon momentum,
$d^{3}p_{\mu}=p_{\mu}^{2}\,dp_{\mu}\,d\Omega_{\mu}$. The
angular integration can be done by symmetry, yielding $4\pi$; then the $\delta$-function eliminates the $dp_{\mu}$ integration, fixing $p_{\mu}$ at
its physical value. The energy $\delta$-function also introduces a factor
$\left|\frac{d}{dp_{\mu}}(m_{\pi}-E_{\mu}-E_{\nu})\right|^{-1}
=|v_{\mu}+v_{\nu}|^{-1}$, where the velocities are evaluated at the fixed momentum
$p_{\mu}$.

The neutrino velocity is simple, $v_{\nu}=1-\frac{4}{3}c_{00}$, but the muon case is
trickier. The general expression for a massive particle's group velocity is
\begin{equation}
\label{eq-v}
v_{j}=
\frac{\pi_{j}}{\sqrt{m^{2}+\vec{\pi}^{2}}}-c_{00}\frac{\pi_{j}}{\sqrt{m^{2}+
\vec{\pi}^{2}}}-2c_{jk}\frac{\pi_{k}}{\sqrt{m^{2}+\vec{\pi}^{2}}}+c_{kl}
\frac{\pi_{j}\pi_{k}\pi_{l}}{(m^{2}+\vec{\pi}^{2})^{3/2}}-2c_{0j},
\end{equation}
which, in the isotropic limit, reduces to
\begin{equation}
v_{\mu}=\frac{p_{\mu}}{\sqrt{m_{\mu}^{2}+p_{\mu}^{2}}}\left\{1+\left[
-\frac{5}{3}+\frac{p_{\mu}^{3}}{3(m_{\mu}^{2}+p_{\mu}^{2})^{3/2}}\right]c_{00}
\right\}.
\end{equation}
Evaluated at the real muon momentum, this becomes
\begin{equation}
\label{eq-vmu}
v_{\mu}=\frac{1-\xi^{2}}{1+\xi^{2}}\left\{1+\left[\frac{2(2+\xi^{2})}{3(1+\xi^{2})}
\left(\frac{1+\xi^{2}}{1-\xi^{2}}-\frac{1-\xi^{2}}{1+\xi^{2}}\right)-\frac{5}{3}
+\frac{1}{3}\left(\frac{1-\xi^{2}}{1+\xi^{2}}\right)^{2}\right]c_{00}\right\},
\end{equation}
which makes $v_{\mu}+v_{\nu}$ the considerably simpler
\begin{equation}
v_{\mu}+v_{\nu}=\frac{2}{1+\xi^{2}}\left[1-\frac{2(2-\xi^{2})}{3(1+\xi^{2})}c_{00}
\right].
\end{equation}

It is now possible to assemble all the pieces of the cross section. Since there is
no Lorentz violation in the meson sector, the kinematic factors associated with
the initial state are not modified. The remaining effects give
\begin{eqnarray}
\label{eq-Gamma-scaling}
\frac{\Gamma}{\Gamma_{0}} & = & \left(\frac{p_{\mu}}{p_{\mu0}}\right)^{2}
\left(\frac{E_{\mu0}}{E_{\mu}}\right)\left(\frac{E_{\nu0}}{E_{\nu}}\right)
\left(\frac{v_{\mu0}+v_{\nu0}}{v_{\mu}+v_{\nu}}\right)
\left|\frac{{\cal M}}{{\cal M}_{0}}\right|^{2} \\
\label{eq-Gamma}
& = & 1+\frac{4}{1-\xi^{2}}c_{00}.
\end{eqnarray}
The first factor on the right-hand-side of (\ref{eq-Gamma-scaling}) arises from the
$p_{\mu}^{2}$ in $d^{3}p_{\mu}$.

The final formula (\ref{eq-Gamma}) is quite simple, and it depends on the same
single quantity $1-\xi^{2}$ that typically characterizes leptonic decay rates.
[The simplicity of the final result also suggests that there may be a more convenient
way of organizing this calculation, in which complicated rational functions of
$\xi^{2}$, such as those in (\ref{eq-vmu}), need not appear.]
Using the physical value of $\xi=0.757$ for pion decay,
$\Gamma=\Gamma_{0}(1+9.4c_{00})$. For heavier mesons, the dependence on $c_{00}$ will
be smaller. Note that $c_{00}$ can also be expressed as
$-\frac{3}{4}(\langle v_{\nu}\rangle-1)$, where $\langle v_{\nu}\rangle$ is the
neutrino speed (averaged over all directions $\hat{v}$) in the rest frame of the
decaying pion.

The modification to $\Gamma$ given
in~(\ref{eq-Gamma}) depends on the mass of the parent particle. Most of the
$c$ dependence in $\Gamma$ comes from the kinematic factors rather than from the
dynamical matrix element. The fact that the kinematics are
more strongly affected when $\xi$ is close to 1 is relatively straightforward to
understand. At small momenta, a small addition to the velocity
produces a much larger fractional
change in the available phase space than would a similar term at large momenta.
This tends to make the decay rate more sensitive to $c$ when the energy available to
the decay products is small.

Since superluminal neutrinos correspond to negative values of $c_{00}$,
superluminal behavior will tend
to reduce the rate at which a pion will decay. This is natural,
since the decay products (both the neutrinos and the corresponding muons)
have dispersion relations for which the energy increases more rapidly than usual
as a function of the three-momentum. Since the energy available
in the decay is fixed to be $m_{\pi}$, the negative value of $c_{00}$ decreases the
momentum that the daughter particles can carry and consequently also decreases the
available phase space.

There are two straightforward kinds of comparisons that might be made,
in order to test the
hypothesis of a nonzero muon neutrino $c$. The first involves a comparison of
the decay rates for pions (or other mesons) moving with different velocities. The
$c_{00}$ in (\ref{eq-Gamma}) is the coefficient in the decaying particle's rest frame.
In terms of the coefficients $c_{TT}$ and $c_{TJ}$ in the sun-centered frame, the
$c_{00}$ is
\begin{equation}
c_{00}=\gamma_{\pi}^{2}\left[c_{TT}+c_{(TJ)}(v_{\pi})_{J}+c_{JK}(v_{\pi})_{J}
(v_{\pi})_{K}\right],
\end{equation}
where $\vec{v}_{\pi}$
is the meson velocity in the sun-centered frame, and $\gamma_{\pi}$ is the
corresponding Lorentz factor $\gamma=(1-v_{\pi}^{2})^{-1/2}$.

Generally, the
size of the Lorentz violation effect grows as the square of the meson energy.
(This novel phenomenon is, of course, in addition to the conventional time dilation
effect that lengthens the apparent lifetime of the meson, as measured by a stationary
observer, by $\gamma_{\pi}$.) Although the absolute magnitude of the pion decay rate
depends on difficult-to-compute hadronic effects, comparisons of the decay rates for
mesons with different
boosts can be used to search for a nonzero $c_{TT}$. Most obviously,
measurements of highly boosted
decays may be compared with the reference value of the decay lifetime for a meson
at rest. Moreover, studying the decays of mesons moving in
different directions adds potential sensitivity to the other coefficients $c_{(TJ)}$
and $c_{JK}$.
Even if, at a given source facility, the mesons produced are traveling in relatively
tight beam, oriented along a fixed compass direction, the rotation of the Earth will
make it possible to sample multiple directions $\hat{v}$. The
current best value of the charged pion lifetime is accurate at the $2\times10^{-4}$
level in measurements with nonrelativistic pions. Experiments of comparable precision
with relativistic pions with $\gamma_{\pi}\gtrsim3$ would be sensitive to anomalies
in the neutrino dispersion relation at the level reported by OPERA. At higher
energies, experiments can be less precise by a factor of $\gamma_{\pi}^{2}$ and
still be capable of confirming the novel effects.

The other way in which this effect might be straightforwardly tested involves
comparisons of multiple meson decay modes. Light charged mesons are generally much
more likely to decay into muon and neutrino pairs than electron and neutrino pairs.
However, the latter decay does occur, and the ratio of the two decay rates can be
predicted quite precisely in the standard model. Tests of lepton universality, which
measure the relative sizes of the branching ratios for decays such as
$\pi^{+}\rightarrow\mu^{+}+\nu_{\mu}$ and $\pi^{+}\rightarrow e^{+}+\nu_{e}$, are
therefore sensitive to the muon $c$ coefficients.

Of course, the $\pi^{+}\rightarrow e^{+}+\nu_{e}$ decay rate is sensitive to
the $c$ coefficients in the electron
sector. However, the coefficients for electrons (and, therefore, their associated
neutrinos) are much more tightly constrained than the coefficients for the
second-generation leptons. The reason is that electrons are stable and extremely
numerous. They play a crucial role in many high-energy phenomena, and their
energy-momentum relation has been mapped out very carefully, even up to PeV
energies. This can be done by studying the radiation emitted by highly boosted
electrons; such electromagnetic emissions are sensitive probes of both electron
velocities and energies. By looking at the ways various processes would be modified
by the inclusion of electron $c$ coefficients and comparing these predictions to
observations of energetic electrons---both in astrophysical
sources~\cite{ref-altschul6} and at particle
accelerators~\cite{ref-altschul22}---all the relevant coefficients may be constrained
at the $10^{-14}$ level or better.


Finally, it is worth noting that for sufficiently large boosts, the decay
$\pi^{+}\rightarrow\mu^{+}+\nu_{\mu}$ may be impossible. If
$E_{\pi}>\sqrt{(m_{\pi}^{2}-m_{\mu}^{2})/2[-c_{TT}-c_{(TJ)}(v_{\pi})_{J}-
c_{JK}(v_{\pi})_{J}(v_{\pi})_{K}]}$, the greater-than-normal
growth of the muon and neutrino energies makes the decay energetically
impossible~\cite{ref-bi,ref-mestres}.
The pion energy $\gamma_{\pi}m_{\pi}$ is insufficient to
produce the highly boosted decay products.
In contrast, in a purely phenomenalistic model, which violates gauge invariance by
giving only the neutrino a modified dispersion relation, there is no threshold above
which the pion is stable~\cite{ref-cowsik}.

If neutrinos do move faster than light, this will have many consequences in particle
physics. Some of these consequences are rather counter-intuitive.
For example, without
also violating electroweak gauge invariance, it is impossible to have
an energy-independent form of Lorentz violation for neutrinos without having it for
the charged leptons as well.
A complete dynamical calculation of the pion decay rate shows that it is fairly
sensitive to this kind of Lorentz violation, and this provides a completely
independent avenue for testing the fascinating OPERA result.


\begin{thebibliography}{99}

\bibitem{ref-opera}T. Adam, {\em et al.} (OPERA collaboration), arXiv:1109.4897.
\bibitem{ref-kost1}D. Colladay, V. A. Kosteleck\'{y}, Phys. Rev. D {\bf 55},
6760 (1997).
\bibitem{ref-kost2}D. Colladay, V. A. Kosteleck\'{y}, Phys. Rev. D {\bf 58},
116002 (1998).
\bibitem{ref-kost4}V. A. Kosteleck\'{y}, C. D. Lane, A. G. M. Pickering,
Phys. Rev. D {\bf 65}, 056006 (2002).
\bibitem{ref-colladay2}D. Colladay, P. McDonald, Phys. Rev. D {\bf 75}, 105002
(2007).
\bibitem{ref-ferrero1}A. Ferrero, B. Altschul, Phys. Rev. D, {\bf 84}, 065030 (2011).
\bibitem{ref-kost3}V. A. Kosteleck\'{y}, R. Lehnert, Phys. Rev. D {\bf 63},
065008 (2001).
\bibitem{ref-tables}V. A. Kosteleck\'{y}, N. Russell, Rev. Mod. Phys. {\bf 83}, 11
(2011).
\bibitem{ref-bi}X.-J. Bi, P.-F. Yin, Z.-H. Yu, Q. Yuan, arXiv:1109.6667.
\bibitem{ref-cohen}A. G. Cohen, S. L. Glashow, Phys. Rev. Lett {\bf 107}, 181803
(2011).
\bibitem{ref-altschul14}B. Altschul, Astropart. Phys. {\bf 28}, 380 (2007).
\bibitem{ref-altschul9}B. Altschul, Phys. Rev. Lett. {\bf 98}, 041603 (2007).
\bibitem{ref-bennett}G. W. Bennett, {\em et al.} (Muon $g-2$ collaboration), Phys.
Rev. Lett. {\bf 100}, 091602 (2008).
\bibitem{ref-carone}C. D. Carone, M. Sher, M. Vanderhaeghen, Phys. Rev. D {\bf 74},
077901 (2006).
\bibitem{ref-kost22}V. A. Kosteleck\'{y}, M. Mewes, Phys. Rev. D {\bf 69}, 016005
(2004).
\bibitem{ref-barger1}V. Barger, J. Liao, D. Marfatia, K. Whisnant, Phys. Rev. D
{\bf 84}, 056014 (2011).
\bibitem{ref-giudice}G. F. Giudice, S. Sibiryakov, A. Strumia, arXiv:1109.5682.
\bibitem{ref-drago}A. Drago, I. Masina, G. Pagliara, R. Tripiccione, arXiv:1109.5917.
\bibitem{ref-longo}M. J. Longo, Phys. Rev. D {\bf 36}, 3276 (1987). 
\bibitem{ref-stodolsky}L. Stodolsky, Phys. Lett. B {\bf 201}, 353 (1988).
\bibitem{ref-bluhm4}R. Bluhm, V. A. Kosteleck\'{y}, C. D. Lane, N. Russell, Phys.
Rev. D {\bf 68}, 125008 (2003).
\bibitem{ref-kost5}D. Colladay, V. A. Kosteleck\'{y}, Phys. Lett. B {\bf 511}
209 (2001).
\bibitem{ref-altschul6}B. Altschul, Phys. Rev. Lett. {\bf 96}, 201101 (2006).
\bibitem{ref-altschul22}B. Altschul, Phys. Rev. D {\bf 82}, 016002 (2010).
\bibitem{ref-mestres}L. Gonzalez-Mestres, arXiv:1109.6630.
\bibitem{ref-cowsik}R. Cowsik, S. Nussinov, U. Sarkar, arXiv:1110.0241.

\end{thebibliography}
\end{document}